\documentclass[fleqn,usenatbib]{mnras}

\usepackage{newtxtext,newtxmath}
\usepackage[T1]{fontenc}
\usepackage{ae,aecompl}

\usepackage{graphicx}	% Including figure files
\usepackage{amsmath}	% Advanced maths commands
\usepackage{amssymb}	% Extra maths symbols

\title[Epoch of HeII reionization]{Constraining the era of helium reionization using fast radio bursts}

\author[M.~Caleb et al.]
{M.~Caleb$^{1}$\thanks{Email: manisha.caleb@manchester.ac.uk},
C.~Flynn$^{2,3}$,
B.W.~Stappers$^{1}$
\\
$^{1}$ Jodrell Bank Centre for Astrophysics, School of Physics and Astronomy, The University of Manchester, Manchester M13 9PL, UK \\
$^{2}$ Centre for Astrophysics and Supercomputing, Swinburne University of Technology, P.O. Box 218, Hawthorn, VIC 3122, Australia\\
$^{3}$ ARC Centre of Excellence for All-sky Astrophysics (CAASTRO)}

\date{Accepted XXX. Received YYY; in original form ZZZ}

\pubyear{2018}

\begin{document}
\label{firstpage}
\pagerange{\pageref{firstpage}--\pageref{lastpage}}
\maketitle
\begin{abstract}
The discovery of fast radio bursts (FRBs) about a decade ago opened up new possibilities for probing the ionization history of the Intergalactic Medium (IGM). In this paper we study the use of FRBs for tracing the epoch of He$\textsc{ii}$ reionization, using simulations of their dispersion measures. We model dispersion measure contributions from the Milky Way, the IGM (homogeneous and inhomogeneous) and a possible host galaxy as a function of redshift and star formation rate. We estimate the number of FRBs required to distinguish between a model of the Universe in which helium reionization occuured at $z = 3$ from a model in which it occurred at $z = 6$ using a 2-sample Kolmogorov-Smirnoff test. We find that if the IGM is homogeneous $\gtrsim 1100$ FRBs are needed and that an inhomogeneous model in which traversal of the FRB pulse through galaxy halos increases the number of FRBs modestly, to $\gtrsim 1600$. We also find that to distinguish between a reionization that occurred at $ z = 3$ or $z = 3.5$ requires $\gtrsim 5700$ FRBs in the range $ 3 \leq z \leq 5$.
\end{abstract}


%%%%%%%%%%%%%%%%%%%%%%%%%%%%%%%%%%%%%%%%%%%%%%%%%%

%%%%%%%%%%%%%%%%% BODY OF PAPER %%%%%%%%%%%%%%%%%%

\section{Introduction}

One of the key questions in present day cosmology is when the reionization of Helium occurred. Following cosmological
recombination at redshift $ z \sim 10^3$, the baryonic gas in the Universe remained primarily neutral. Given that presently
most of the observable baryonic gas is ionized, phase transitions of this gas must have taken place at some redshift $z < 10^3$. The progress of reionization is regulated by the ioinizing radiation escaping the host galaxies from e.g. quasars and stars. 
Reionization typically refers to the events related to neutral hydrogen (H\textsc{i}) and helium (He\textsc{i}) at $z \sim 6$ when 
the atoms lost their first outer electrons and marked an important change in the structure of the Universe known as the Epoch of Reionization \citep[EoR;][]{Fan, Zaroubi, Singh}. Present observational constraints on the EoR have mostly been indirect but direct observations of the redshifted 21-cm signal from neutral hydrogen using present and upcoming experiments such as the Low-Frequency Array \citep[LOFAR;][]{vanHaarlem}, Murchison Wide-field array \citep[MWA;][]{Tingay}, Precision Array for Probing the Epoch of Reionization \citep[PAPER;][]{Parsons} and the Hydrogen Epoch of Reionization Array \citep[HERA;][]{DeBoer} are expected to provide stronger constraints. 

Following the EoR, the high ionization potential (54.4 eV) of singly ionised helium (He\textsc{ii}) prevented it from being further reionized  until a sufficient build-up of hard-spectra quasars provided enough photons to complete the reionization. He\textsc{ii} is 
expected to have undergone a second reionization at $ z \sim 3$ with significant contributions from stars and quasars leaving an impression on the IGM. The strongest evidence of this is seen through observational signatures in the far ultraviolet spectra of the He\textsc{ii} Ly$\alpha$ forest along the lines-of-sight to several quasars at $z \sim 3$ \citep[e.g.][]{SyphersHeII2, SyphersHeII}. The Ly$\alpha$ transition for neutral hydrogen, H\textsc{i} exhibits as absorption features in 
the spectra of distant quasars just as it does for the transition of singly ionized helium, He\textsc{ii}. However observations of the He\textsc{ii} Ly$\alpha$ forest for the second reionization is mostly obscured by contamination due to the relative abundance of high-density systems at low redshifts \citep[see][and references therein]{Plante}. The comparatively low number of lines-of-sight that show the Ly$\alpha$ forest signature for He\textsc{ii} leaves much statistical uncertainly about the exact timing and nature of the reionization process. Other evidence of reionization is the temperature evolution of the intergalactic medium (IGM) with a quick change at $ z \sim 3$ \citep{McQuinnHeII}. The comparatively lower redshift of He\textsc{ii} reionization makes it more accessible to observations.

% Several numerical simulations
% \citep{McQuinnHeII, Puchwein} and observations \citep{Worseck} have been performed over the years to improve our understanding of the total optical depth to 
% reionization of the cosmic microwave background (CMB), $\tau_\mathrm{CMB}$ \citep{FialkovLoeb} and for studying the structure and thermal evolution
% of the intergalactic medium \cite[IGM;][]{Plante}. The Ly$\alpha$ transition for neutral hydrogen, H\textsc{i} exhibits as absorption features in 
% the spectra of distant quasars just as it does for the transition of singly ionized helium, He\textsc{ii}. However observations of the He\textsc{ii} Ly$\alpha$ forest for the second reionization is mostly obscured by contamination due to the relative abundance of high-density systems at low redshifts \citep{Plante}.

FRBs are
intense bursts ($\sim$ Jy) of coherent emission of unknown origin, lasting only a few milliseconds. Over the last decade or so, realization has grown that this and other properties could be used as a cosmological probe (see \citet{Keane2018} for a recent review). In particular, their short durations enable direct
measurements of the integrated column densities of ionized plasma known as dispersion measures (DMs), along the observed
lines-of-sight thereby tracing all the ionized baryons along its propagation path. Electron density models of the Milky Way \citep[e.g.][]{Cordes, Yao} are unable to account for more than a few percent of the total observed DM values, 
thereby suggesting extragalactic/cosmological origins for FRBs. Precise localization to a host galaxy is however
necessary to determine the true extragalactic  or cosmological nature of FRBs, as was demonstrated for 
FRB 121102 \citep{Chatterjee, SriHarsh}.
% Their use as cosmological probes is also highly dependent of their localisation to a host galaxy. 
FRBs with redshifts obtained from host galaxies could be used to probe the epoch of helium reionization \citep{Deng, Zheng}, map the 
intergalactic magnetic fields \citep{JPska}, locate the `missing' baryons in the Univers \citep{McQuinn} and 
study the dark energy equation-of-state \citep{Zhou}.

In this letter we study the possibility of probing the epoch of He\textsc{ii} reionization  with FRBs through simulations of their observed DMs and estimate the numbers of FRBs required, assuming they have been localised to a host galaxy with a measured redshift. We present our DM model for the simulated FRBs in Section \ref{sec:model} followed by our analyses and results in Section \ref{sec:results}. We finally present our discussion and conclusions in Section \ref{sec:conc}.

\section{Modeling dispersion measures}
\label{sec:model}

The DM of an FRB parameterises the delay in arrival time as a function of frequency of a pulse which is given by,

\begin{equation}
\label{eq:tdelay}
    \Delta t = 4.15 \, \mathrm{ms} \, \times  \Bigg[ \bigg( \frac{\nu_{1}}{\mathrm{GHz}}\bigg)^{-2} - \bigg( \frac{\nu_{2}}{\mathrm{GHz}}\bigg)^{-2} \Bigg] \, \times \Bigg(\frac{\mathrm{DM}}{\mathrm{pc \, cm^{-3}}} \Bigg)
\end{equation}

\noindent and is an observable quantity. This delay in arrival time, in the observer's frame is $\Delta t = \Delta t' (1+z)$, where $\Delta t'$ is the rest-frame delay. Similarly, the observed frequencies in the observer's frame are $\nu_{1} = \nu_{1}'(1+z)$ and $\nu_{2} = \nu_{2}'(1+z)$ with $\nu_{1}'$ and $\nu_{2}'$ being the source rest-frame frequencies.
Using these relations, the DM measured in the observer's frame in Equation \ref{eq:tdelay} is related to the free electron density along the line-of-sight as,

\begin{equation}
    % \mathrm{DM} = \int^{z}_{0} n_\mathrm{e} \, dl.
    \mathrm{DM} = \int^{z}_{0} \frac{n_\mathrm{e}}{(1+z)} \, dl.
\end{equation}

% \noindent where $n_\mathrm{e}$ is the free electron density along the line-of-sight and $dl$ is the path length integral. 
% {\color{red} Uncertainty in DM$_\mathrm{obs}$ is negligible. The average of all DMs at 1.4 GHz.}
\noindent For an FRB originating at cosmological distances, the observed DM value can be expressed as the sum of several contributions along the path traversed given by,

\begin{equation}
\label{eq:dm}
    \mathrm{DM_{obs}} = \mathrm{DM_{MW}} + \mathrm{DM_{IGM}} + \frac{\mathrm{DM_{host} + DM_{source}}}{(1+z)}
\end{equation}

\noindent where $\mathrm{DM_{MW}}$ is the contribution from the Milky Way (MW), $\mathrm{DM_{IGM}}$ is the contribution from the 
diffuse homogeneous/inhomogeneous IGM along the line-of-sight, $\mathrm{DM_{host}}$ is any contribution from the interstellar medium 
(ISM) of the host galaxy and $\mathrm{DM_{source}}$ is the contribution from the immediate vicinity of the FRB progenitor.
The lack of understanding of the physical origins of FRBs, hinders our ability to ascertain the $\mathrm{DM_{host}}$ and $\mathrm{DM_{source}}$ components of the total observed DM reliably.
However due to the cosmological redshifting of frequency, their combined contribution is diluted by a factor of $(1+z)$ to the Earth observer from the rest-frame observer \citep{Ioka, Deng} and is likely to be small. As a result the $\mathrm{DM_{IGM}}$ is the dominant contribution to the 
observed DM as the number density of free electrons scales as $(1+z)^3$.
% The contribution from the MW can be easily accounted for using existing models as described in Section \ref{sec:DMgal}.

% \subsection{DM$_\mathrm{MW}$}
\subsection{DM contribution from the Milky Way}
\label{sec:DMgal}

The MW component of the total observed DM can be reasonably estimated 
using either the NE2001 \citep{Cordes} or the YMW16 model \citep{Yao} and removed from the data with modest uncertainties. 
Both models use Galactic pulsars to map the 
the integrated electron density along any given line-of-sight through the MW which strongly decreases as a function
of Galactic latitude $|b|$ from $\sim 10^3$ pc cm$^{-3}$ near the Galactic centre to an average of $\sim 100$ pc cm$^{-3}$
at  $10^{\circ} \leq |b| \leq 40^{\circ}$. It should be noted that the estimate of DM at $|b| > 40^{\circ}$ are less reliable
due to the lack of pulsars in the Galactic halo. Since the electron density content in the Galactic halo is relatively 
low with a correspondingly small DM contribution ($\sim 30$ pc cm$^{-3}$) as seen from simulations by \cite{Dolag}, 
we do not include a term for the halo contribution in Equation \ref{eq:dm}. 
% Overall, since the observed population of  FRBs have relatively high latitudes, the $\mathrm{DM_{MW}}$ contribution is relatively small.
Overall, for an extragalactic source like an FRB, the contribution from the Milky Way is relatively small over most of the sky, and can be avoided in FRB searches if needed.
% over the vast majority of the sky as viewed from earth.

% \subsection{DM$_\mathrm{IGM}$}
\subsection{DM contribution from IGM}
\label{sec:DMigm}

The free electron density at any given redshift can be expressed as \citep{Deng},

\begin{equation}
\label{eq:electrondensity}
n_{e} = \frac{\rho_{c,0} \Omega_\mathrm{b}f_\mathrm{IGM}f_{e}(z)}{m_{p}}\, \times (1+z)^3,
\end{equation}

\noindent where $\rho_{c,0}$ is the critical mass density, $\Omega_{b} = 0.049$ is the present baryon density of the Universe \citep{Planck} and,

\begin{equation}
    f_{e}(z) = \frac{3}{4} \upchi_{e,\mathrm{H}}(z) + \frac{1}{8} [\upchi_{e,\mathrm{He\textsc{ii}}}(z) + \upchi_{e,\mathrm{He\textsc{iii}}}(z)],
\end{equation}

\noindent with the terms $\upchi_{e,\mathrm{H}}(z)$ and
$\upchi_{e,\mathrm{He}}(z)$ representing the ionization mass fractions of hydrogen and helium respectively. Strictly speaking, the baryon mass fraction in the IGM, $f_\mathrm{IGM}$ varies slightly a function of redshift from 0.9 at $z \gtrsim 1.5$ to 0.82 at $z \leq 0.4$ \citep{Shull}. To first order, we assume it to be a constant at 0.83 \citep{Fukugita1998, Shull}. For a path length $dl$ given by \citep{Deng},

\begin{equation}
\label{eq:pathlength}
    dl = \frac{1}{1+z} \, \frac{c}{H_{0}} \, \frac{dz}{\sqrt{\Omega_{m}(1+z)^3 + \Omega_{\Lambda}}},
\end{equation}

\noindent where the matter density $\Omega_{m} = 0.308$, the vacuum density 
$\Omega_{\Lambda} = 0.6911$ and Hubble constant $H_{0} = 67.74$ km s$^{-1}$ Mpc$^{-1}$ \citep{Planck}, we can combine Equations \ref{eq:electrondensity} and \ref{eq:pathlength} to obtain a relation for the homogeneous IGM contribution to the DM related to the FRB redshift ($z$) as \citep{Zheng, Deng},

\begin{equation}
\label{eq:DMigm}
\mathrm{DM_{IGM}} (z) = \frac{3cH_{0}\Omega_\mathrm{b}f_\mathrm{IGM}}{8\pi G m_{p}} \, \times \, \int_{0}^{z} \frac{f_{e}(z) \, (1+z) \, dz}{\sqrt{\Omega_{m}(1+z)^3 + \Omega_{\Lambda}}}
\end{equation}

\noindent under the assumption of a flat Universe.
% $n_\mathrm{e} = (3H_{0}^2 c \Omega_\mathrm{b}/8\pi G m_\mathrm{p})(1+z)^3$ and $dz/dl = (1+z)H_{0}\sqrt{\Omega_{m}(1+z)^3 + \Omega_{\Lambda}}$ 

% \begin{equation}
% f_{e}(z) = \frac{3}{4} \upchi_{e,\mathrm{H}}(z) + \frac{1}{8} [\upchi_{e,\mathrm{He\textsc{ii}}}(z) + \upchi_{e,\mathrm{He\textsc{iii}}}(z)].
% \end{equation}
We assume that hydrogen is fully ionized at $z = 6$ (i.e. $\upchi_{e,\mathrm{H}} = 1$ and  $\upchi_{e,\mathrm{He\textsc{ii}}} = 1$) 
and that the epoch of HeII reionization occurs at $z = 3$ in sharp transition (i.e. $\upchi_{e,\mathrm{He}} = 1$, 
$\upchi_{e,\mathrm{He\textsc{ii}}} = 0$ and  $\upchi_{e,\mathrm{He\textsc{iii}}} = 1$) \citep{Sokasian, Gao, Plante}. It is important to note that we assume the transition to have  occurred at the same time across the entire Universe.
% since the helium epoch of reionization is estimated to occur between $3 \leq z \leq 4$ \citep{Sokasian}.

\begin{figure*}
\centering
\includegraphics[width=6.5 in]{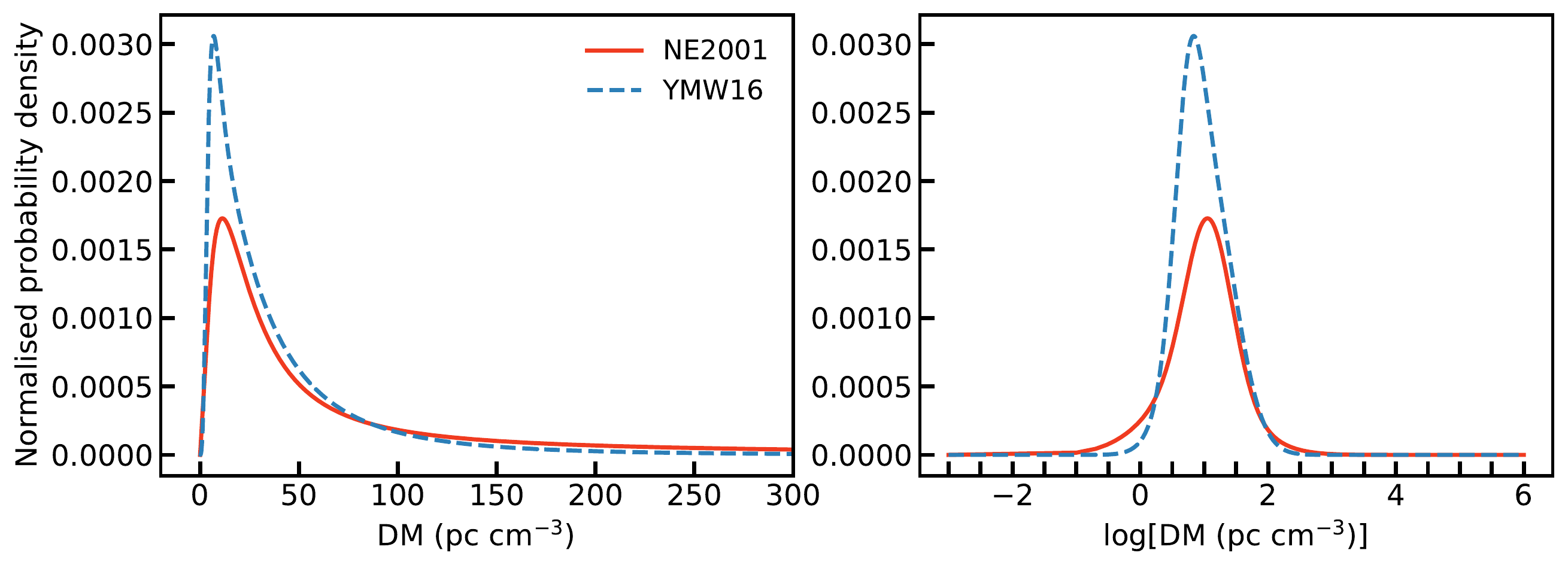}
\caption{Normalised probability density distributions of the `all galaxies' sample from \citep{Luo} for 
the NE2001 and YMW16 electron density maps in linear (left panel) and log scales (right panel). 
The two curves have been normalised to equal areas.}
\label{fig:pdfs}
\end{figure*}

% \subsection{DM$_\mathrm{host}$}
\subsection{DM contribution from ISM in the host galaxy}
\label{sec:dmhost}

Knowledge of the DM contribution from the host galaxy is essential to disentangle the amount contributed by the IGM from the total
observed DM.
\cite{Luo} model the probability distributions of DM contributions for early-type galaxies (ETGs), late-type 
galaxies (LTGs) and a combination of both types of galaxies called all galaxies (ALGs).
The MW and M87 galaxies are chosen to be templates for LTGs and ETGs respectively since their
electron density profiles and H$\alpha$ luminosities are well studied \citep{Cavagnolo, Cordes, Yao}. For each of the template 
galaxies, $10^6$ FRBs are simulated at randomly chosen positions following a Galactic stellar distribution \citep[e.g.][]{Young}. A detailed
modelling of the stellar distribution and the gas fractions can be found in \cite{Luo}. The FRBs in their model are generated 
over the full-sky with various line-of-sight inclination angles of their host galaxies using a Monte Carlo method, and the DM probability density function (PDF) 
is calculated by integrating the electron density along the chosen line-of-sight out to the edge of the galaxy. An important assumption made by \cite{Luo} is that all galaxies, dwarfs and giants alike, are equally likely to host FRBs due to the uncertainty in their origins. A consequence of this assumption is that low mass galaxies are more likely to be FRB hosts (as is the case for the repeater FRB 121102), due to simply to their much higher numbers. The zero-redshift rest-frame DM distribution function of the simulated ETGs and LTGs (their Figure 3) are calculated by scaling a H$\alpha$ luminosity from a luminosity function and an effective galaxy radius from an $r$-band luminosity function 
% along with a randomly chosen DM value from the template PDF, 
to the H$\alpha$ luminosity and $r$-band luminosity of the template. The resulting PDF of the 
sample of ALGs is similar to that of the LTGs due to their dominance (76.3$\%$) in the sample compared to the
ETGs (23.7$\%$) \citep{Luo}. 

\cite{Luo}
present two samples of LTGs modelled after the NE2001 \citep{Cordes} and YMW16 \citep{Yao} Galactic electron density 
maps from their Monte Carlo method. As a result we adopt the ensemble zero-redshift rest-frame DM PDFs of both ALG samples (NE2001 and YMW16) shown in Figure \ref{fig:pdfs} in our simulations. Galaxies whose electron density distribution follows that
of the NE2001 model are seen to produce $\sim$14 times more FRBs with DMs greater than the 3$\sigma$ value 
of the YMW16 model (see Figure \ref{fig:pdfs}). This is supported by the fact that unlike the NE2001 model the YMW16 model includes no
over-dense regions, thereby yielding a much lower average electron density \citep{Yao}.

\begin{figure*}
\centering
\includegraphics[width=6.0 in]{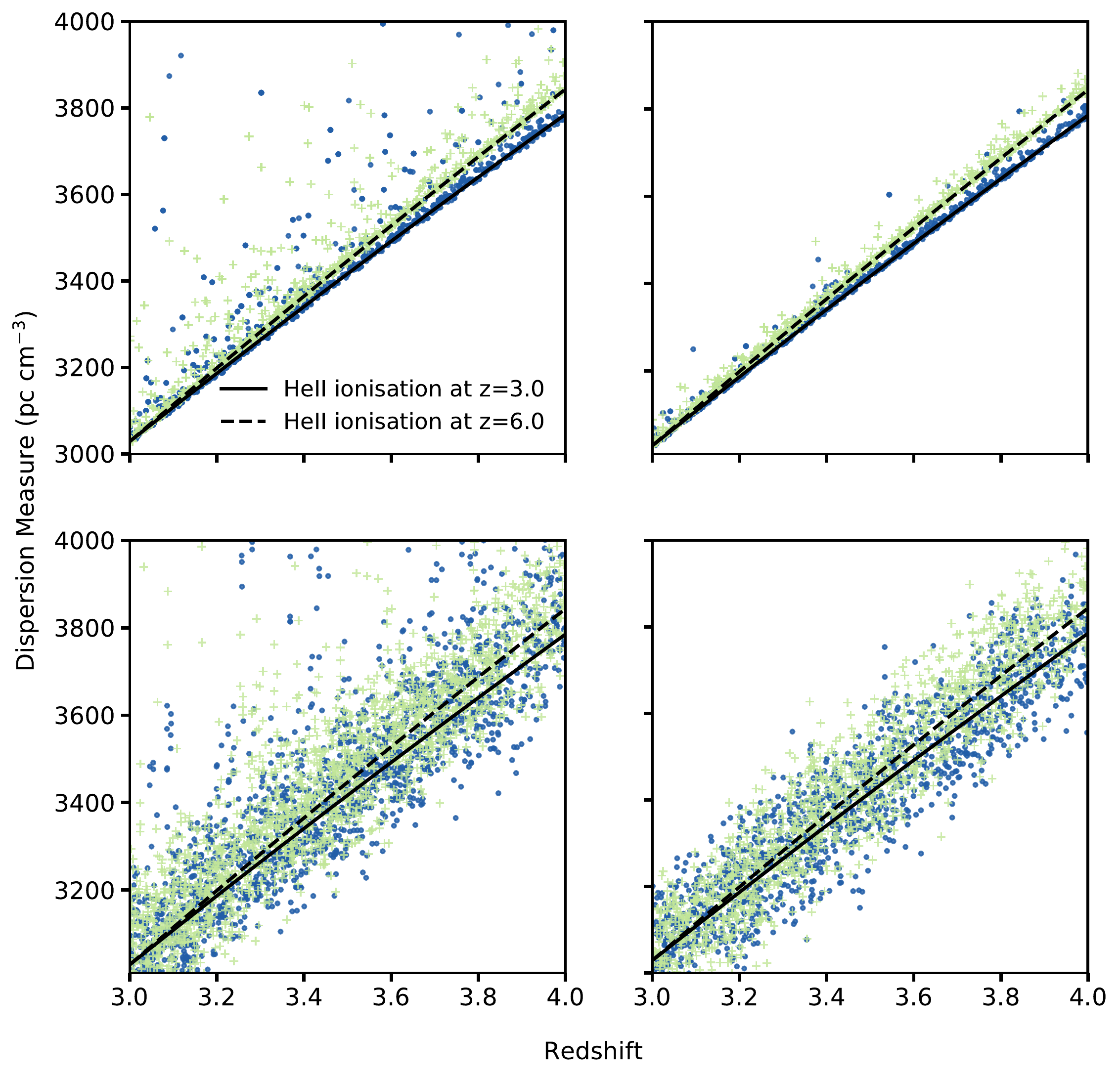}
\caption{Distribution of FRB DMs as a function of redshift. The dashed and solid lines represent 
two different model Universes in which the He\textsc{ii} reionization occurred at $z = 3$ and one in which it occurred at $z = 6$. The squares and dots represent the MW and host corrected observed DM values for each of the models (solid and dashed)
from Equation \ref{eq:dm}. The top row panels (\textit{top left}: Ne2001, \textit{top right}: YMW16) represent FRBs propagating through a homogeneous IGM and the bottom row panels (\textit{bottom left}: NE2001, \textit{bottom right}: YMW16)  represent propagation through an inhomogeneous IGM. The DM due to the local environment surrounding the progenitor has not been modelled.}
\label{fig:HeII}
\end{figure*}

% \subsection{DM$_\mathrm{source}$}
\subsection{DM contribution from progenitor environment}

Despite the association of FRB 121102 with a dwarf galaxy \citep{Chatterjee, SriHarsh, Marcote} the nature of the 
source producing the FRB is still under debate. Compact objects, particularly neutron star models at cosmological 
distances, are favoured as FRB progenitors due to their ability to account for the millisecond timescales and observed polarization features. In this case, sources embedded in pulsar wind nebulae or supernova
remnants, or near H\textsc{ii} regions will be associated with high electron densities. But the DM$_\mathrm{source}$ contribution
is restricted by the condition that the plasma frequency should not exceed the radiation frequency thereby allowing the
free propagation of radiation. This limit on the density of the region local to the source along with the cosmological
frequency-shift for the range of redshifts of interest in these simulations make the contributions likely to be small. 
Due to the large uncertainty in the progenitor we do not factor in a value for DM$_\mathrm{source}$.

\section{Analyses and Results}
\label{sec:results}

In our simulations we model two different Universes, one in which He\textsc{ii} reionization occurs in sharp transition at 
$z = 3$ and the other in which it did at $z = 6$, represented respectively by the solid
and dashed lines in Figure \ref{fig:HeII}. Assuming that FRBs are detected out to $z \sim 6$, we draw 
$N_\mathrm{FRBs}$ from a Gaussian redshift distribution centred on $z = 3$ with a $1\sigma$ spread of $z = 1$ fassuming a 
flat $\Lambda$CDM cosmology with the parameters given earlier in Section \ref{sec:DMigm}. The Gaussian is intended to represent the typical distribution of DMs one gets in practice in a survey at a given telescope. A Gaussian is not a very good approximation to the shape of the DM distribution -- it typically has a longer tail to high DM than a Gaussian \citep{Xu, Caleb_sim, CWalker} - but a full blown model in which FRBs are simulated with all the selection effects included out to $z = 10$, with unknown certain assumptions needing to be made about the numbers (or even the existence) of FRBs at higher $z$ is beyond the scope of this paper. The overall Gaussian shape of the DM distribution is affected by (1) the sensitivity of the survey -- higher sensitivity yields higher DM events (2) the frequency and time resolution of the survey -- higher frequency resolution allows FRBs to be probed to higher DMs (3) the area of the survey -- larger area surveys catch more low DM FRBs (presently, as they typically have lower sensitivity as well, but this is changing rapidly with the CHIME \citep{Bandura} and MeerKAT \citep{meertrap} coming on-line). As we aim to keep our model simple, we use a Gaussian distribution. Each FRB at its chosen redshift is assigned a corresponding DM$_\mathrm{IGM}(z)$ from Equation \ref{eq:DMigm}.  We draw a random host galaxy DM from the zero-redshift rest-frame DM PDF 
and scale it to the redshift of the simulated FRB based on the star formation rate at that redshift \citep{Hopkins} 
given by,

\begin{equation}
  \dot{\rho}_{*}(z) =  \frac{0.017 + 0.13z}{1+(z/3.3)^{5.3}} \, M_\odot \, \mathrm{yr}^{-1} \, \mathrm{Mpc}^{-3}.
\end{equation}

Figure \ref{fig:HeII} shows the results of our simulations for $3 \leq z \leq 4$. As detailed in Section \ref{sec:dmhost}, FRBs following the NE2001 model electron density distribution exhibit a larger scatter in their DMs.
% We perform a 2-sample Kolmogorov-Smirnoff (KS) test between the two FRB distributions in each case of chosen electron density model to determine the number required to distinguish between them. 
For each of the host DM models (NE2001, YMW16) we perform a 2-sample Kolmogorov-Smirnoff (KS) test to determine whether we can distinguish between FRBs with DM$_\mathrm{IGM}$ contributions when He$\textsc{ii}$ reionization happened at $z = 3$ and $z = 6$. 
A $p$-value of $<0.05$ is our 
criterion for deciding if the two distributions of FRBs differ. If the Galactic contribution is drawn from the YMW16 model, we require $\sim 950$ FRB
events in the range $3 \leq z \leq 4$ (see Figure \ref{fig:HeII}) with measured redshifts to distinguish between a Universe in which He\textsc{ii}
reionization that occurred at $z = 3$ and one in which it occurred at $z = 6$ at the 95$\%$ confidence level. For
the NE2001 model we require $\sim 1300$ FRB events to distinguish between our two simulated models at the same confidence level due to the comparatively larger scatter in host DMs. We emphasize that all these galaxies are required to have measured redshifts and reasonably good corrections ($\sim 50 \%$) for the host galaxy DM.

We have so far only considered the effect and contribution from a diffuse homogeneous IGM on the observed DM. This however is likely to be too simple a model, due to the presence of free electrons in the halos of potential intervening galaxies. 
\cite{McQuinn} simulate the sightline-to-sightline variation in the observed DMs for different galaxy halo mass models (their Figure 1) and estimate the standard deviation in mean value of DM$_\mathrm{IGM}$ to be 180 - 400 pc cm$^{-3}$ in the range $0.1 \leq z \leq 1.5$. The halo model in which the distribution of baryons traces dark matter results in the largest dispersion of 400 pc cm$^{-3}$ from the mean expected value, at $z = 1$. Since the curve appears to plateau at $z > 1.2$, we model the fluctuations in DM$_\mathrm{IGM}$ due to the halos of intervening galaxies as a Gaussian centred on the mean value of DM$_\mathrm{IGM}$ at that redshift with a 3$\sigma$ spread of 400 pc cm$^{-3}$. The results are shown in Figure \ref{fig:HeII}. We once again perform a 2-sample KS-test between the two distributions. We require at least $\sim 1350$ (YMW16) and $\sim 1860$ (NE2001) FRB events with redshift associations to distinguish between the two simulated models of the Universe at the 95\% confidence level. We also consider how well we could constrain the epoch of He$\textsc{ii}$ reionization. We estimate that for a given sample of FRBs between $3 \leq z \leq 5$, $\sim 5700$ FRBs with measured redshifts are required to distinguish between a He$\textsc{ii}$ reionization that took place at $z = 3$ and $z = 3.5$.

% \begin{figure*}
% \centering
% \includegraphics[width=6.0 in]{Fig3.pdf}
% \caption{Distribution of FRB host galaxy DMs as a function of redshift. The dashed and solid lines represent 
% two different model Universes in which the HeII reionization occurred at $z = 3$ and $z = 6$ respectively.
% The squares and dots represent the MW corrected observed DM values for each of the models (solid and dashed)
% from Equation \ref{eq:dm}. An error in the DM from the IGM is modelled due to potential intervening galaxies. The DM due to the local environment surrounding the progenitor has not been modelled.}
% \label{fig:HeII_IGM}
% \end{figure*}

\section{Discussion and Conclusion}
\label{sec:conc}

In this paper we model the observed DMs of FRBs as a function of redshift to determine if we can measure the epoch of He\textsc{ii} reionization purely based on the DMs. Our first simulations are based on a simple model for the IGM which assumes a homogeneous electron density distribution 
at all redshifts. The host galaxies of the FRBs have DMs modelled using the results of \cite{Luo}. We ignore the DM contribution local to the FRB's environment (i.e. surrounding the source) due to the as yet quite uncertain nature of the progenitors and due to the likely dilution of its effect by the $(1+z)$ cosmological frequency-shift. In such an uncomplicated scenario we find we require of order 1100 FRBs with measured redshifts and DMs in the range $3 \leq z \leq 4$ to distinguish a Universe in which He$\textsc{ii}$ reionization occured at $z = 3$ from one in which it occurred at $z = 6$.

According to \cite{McQuinn} the values of $\mathrm{DM_{IGM}}$ for two different FRBs at similar
redshifts could vary considerably depending on the inhomogeneities along the line-of-sight such as the halos of intervening
galaxies, thereby making the case more complicated. The standard deviation around the mean DM$_\mathrm{IGM}$(z) is 180 - 400 pc cm$^{-3}$ for
$0.5 < z < 1$ depending on the baryon distribution in the different Galactic halo models following different gas profiles \citep{McQuinn}. We adopt the sightline dependent variation in DM for the galaxy halo model in which the baryons trace the dark matter halo profile above a threshold of $10^{10} M_\odot$. We model the evolution of the DM contribution from the halos of intervening galaxies with redshift, as a normal distribution with a 3$\sigma$ scatter of 400 pc cm$^{-3}$. Even in such an extreme case with uncertainties in both DM$_\mathrm{IGM}$ and DM$_\mathrm{host}$ we are still able to distinguish between our two models of the Universe. However, compared to a homogeneous IGM, in our simulations of an imhomogeneous IGM we need at least $\sim 1600$ FRBs with associated redshifts to determine when the epoch of reionization occurred. We also estimate of order $\sim 5700$ FRBs are needed in the range $3 \leq z \leq 5$ to distinguish between a He$\textsc{ii}$ that took place at $z = 3$ and $z = 3.5$ at the 95\% confidence level.

The redshifts associated with FRBs require localisations better than a few arcsec of the FRB themselves, or possibly via radio afterglows, if they occur. High spatial resolution is required to localise sufficiently well to perform optical and other multi-wavelength follow-ups in order to measure the redshift of the host galaxy. With radio interferometers like the Australian Square Kilometre Array Pathfinder \citep[ASKAP;][]{Bannister} and UTMOST \citep{BailesUTMOST} currently doing such FRB searches, and MeerKAT \citep{meertrap} coming online, it is not unreasonable to expect thousands of FRBs to have sufficiently accurate localisations within a few years. 

Major caveats pertain to following up such large numbers of FRBs. To probe redshifts beyond $z \approx 3$, FRBs with DMs $\gtrsim 3000$ need to be found, which requires sensitive instruments with high frequency resolution in order to avoid the effects of dispersion smearing (only one FRB is currently known with a DM of more than 2000 pc\,cm$^{-3}$). \cite{BZhang} has shown that high sensitivity telescopes with large apertures (e.g. FAST) are capable of finding FRBs out to redshifts as high as $z \sim 15$, assuming their progenitors actually form at such early times. Follow-up of thousands of localised host galaxies for $z \gtrsim 3$ requires high localisation accuracy, due to the source density on the sky \citep{Conselice, Eftekhari2017} and is strongly dependent on the typical host galaxies for FRBs, which remains to be determined. While spectroscopic redshifts would be ideal, we note that photometrically obtained redshifts could be of sufficient quality : photometric redshifts show a scatter $dz/(1+z)$ \citep{Straatman} of only $\approx 0.03$ \citep{Ilbert} around the spectroscopic-photometric redshift relation, which corresponds to a an additional scatter of order 30 pc\,cm$^{-3}$ in the redshift-DM relation, smaller than the scatter in our modeling (due to the host galaxy properties and/or inhomegeneities in the IGM). Despite these caveats, the combination of next generation radio telescopes like the Square Kilometer Array\footnote{https://www.skatelescope.org/} and optical telescopes like the Large Synoptic Survey Telescope\footnote{https://www.lsst.org/} and the James Webb Space Telescope\footnote{https://www.jwst.nasa.gov/} could transform our understanding of the host galaxy DM contributions and our ability to study the epoch of He\textsc{ii} reionization.  

% 1) If we simulate FRBs within small dz values the average of the DMs should approach the model?
% 2) Inclination angles and localisation

% 1) peak at zero redshift DM host
% 2) multiply the zero redshift DM host with SFr at that z

% 3) Zhou - at least 100 FRBs per bin in case of imhomogeneous IGM 

% 4) This needs for us to simulate 5000 FRBs

% 5) But in the case of homogeneous IGM lesser number of FRBs can be simulated

\section*{Acknowledgements}
MC would like to thank Rui Luo, Kejia Lee, Charles Walker and Themiya Nanayakkara for useful discussions. MC and BWS acknowledge funding from the European Research Council (ERC) under the European Union's Horizon 2020 research and innovation programme (grant agreement No 694745). CF acknowledges financial support by the Beckwith Trust.

%%%%%%%%%%%%%%%%%%%%%%%%%%%%%%%%%%%%%%%%%%%%%%%%%%



\bibliographystyle{mnras}
\bibliography{refs} % if your bibtex file is called example.bib

\begin{thebibliography}{}
\makeatletter
\relax
\def\mn@urlcharsother{\let\do\@makeother \do\$\do\&\do\#\do\^\do\_\do\%\do\~}
\def\mn@doi{\begingroup\mn@urlcharsother \@ifnextchar [ {\mn@doi@}
  {\mn@doi@[]}}
\def\mn@doi@[#1]#2{\def\@tempa{#1}\ifx\@tempa\@empty \href
  {http://dx.doi.org/#2} {doi:#2}\else \href {http://dx.doi.org/#2} {#1}\fi
  \endgroup}
\def\mn@eprint#1#2{\mn@eprint@#1:#2::\@nil}
\def\mn@eprint@arXiv#1{\href {http://arxiv.org/abs/#1} {{\tt arXiv:#1}}}
\def\mn@eprint@dblp#1{\href {http://dblp.uni-trier.de/rec/bibtex/#1.xml}
  {dblp:#1}}
\def\mn@eprint@#1:#2:#3:#4\@nil{\def\@tempa {#1}\def\@tempb {#2}\def\@tempc
  {#3}\ifx \@tempc \@empty \let \@tempc \@tempb \let \@tempb \@tempa \fi \ifx
  \@tempb \@empty \def\@tempb {arXiv}\fi \@ifundefined
  {mn@eprint@\@tempb}{\@tempb:\@tempc}{\expandafter \expandafter \csname
  mn@eprint@\@tempb\endcsname \expandafter{\@tempc}}}

\bibitem[\protect\citeauthoryear{{Bailes} et~al.,}{{Bailes}
  et~al.}{2017}]{BailesUTMOST}
{Bailes} M.,  et~al., 2017, \mn@doi [\pasa] {10.1017/pasa.2017.39}, \href
  {http://adsabs.harvard.edu/abs/2017PASA...34...45B} {34, e045}

\bibitem[\protect\citeauthoryear{{Bandura} et~al.,}{{Bandura}
  et~al.}{2014}]{Bandura}
{Bandura} K.,  et~al., 2014, in Ground-based and Airborne Telescopes V. p.
  914522 (\mn@eprint {arXiv} {1406.2288}), \mn@doi{10.1117/12.2054950}

\bibitem[\protect\citeauthoryear{{Bannister} et~al.,}{{Bannister}
  et~al.}{2017}]{Bannister}
{Bannister} K.~W.,  et~al., 2017, \mn@doi [\apjl] {10.3847/2041-8213/aa71ff},
  \href {http://adsabs.harvard.edu/abs/2017ApJ...841L..12B} {841, L12}

\bibitem[\protect\citeauthoryear{{Caleb}, {Flynn}, {Bailes}, {Barr},
  {Hunstead}, {Keane}, {Ravi}  \& {van Straten}}{{Caleb}
  et~al.}{2016}]{Caleb_sim}
{Caleb} M.,  {Flynn} C.,  {Bailes} M.,  {Barr} E.~D.,  {Hunstead} R.~W.,
  {Keane} E.~F.,  {Ravi} V.,   {van Straten} W.,  2016, \mn@doi [\mnras]
  {10.1093/mnras/stw175}, \href
  {http://adsabs.harvard.edu/abs/2016MNRAS.458..708C} {458, 708}

\bibitem[\protect\citeauthoryear{{Cavagnolo}, {Donahue}, {Voit}  \&
  {Sun}}{{Cavagnolo} et~al.}{2009}]{Cavagnolo}
{Cavagnolo} K.~W.,  {Donahue} M.,  {Voit} G.~M.,   {Sun} M.,  2009, \mn@doi
  [\apjs] {10.1088/0067-0049/182/1/12}, \href
  {http://adsabs.harvard.edu/abs/2009ApJS..182...12C} {182, 12}

\bibitem[\protect\citeauthoryear{{Chatterjee} et~al.,}{{Chatterjee}
  et~al.}{2017}]{Chatterjee}
{Chatterjee} S.,  et~al., 2017, \mn@doi [\nat] {10.1038/nature20797}, \href
  {http://adsabs.harvard.edu/abs/2017Natur.541...58C} {541, 58}

\bibitem[\protect\citeauthoryear{{Conselice}, {Wilkinson}, {Duncan}  \&
  {Mortlock}}{{Conselice} et~al.}{2016}]{Conselice}
{Conselice} C.~J.,  {Wilkinson} A.,  {Duncan} K.,   {Mortlock} A.,  2016,
  \mn@doi [\apj] {10.3847/0004-637X/830/2/83}, \href
  {http://adsabs.harvard.edu/abs/2016ApJ...830...83C} {830, 83}

\bibitem[\protect\citeauthoryear{{Cordes} \& {Lazio}}{{Cordes} \&
  {Lazio}}{2003}]{Cordes}
{Cordes} J.~M.,  {Lazio} T.~J.~W.,  2003, preprint, \href
  {http://adsabs.harvard.edu/abs/2003astro.ph..1598C} {} (\mn@eprint {}
  {0207156})

\bibitem[\protect\citeauthoryear{{DeBoer} et~al.,}{{DeBoer}
  et~al.}{2017}]{DeBoer}
{DeBoer} D.~R.,  et~al., 2017, \mn@doi [\pasp]
  {10.1088/1538-3873/129/974/045001}, \href
  {http://adsabs.harvard.edu/abs/2017PASP..129d5001D} {129, 045001}

\bibitem[\protect\citeauthoryear{{Deng} \& {Zhang}}{{Deng} \&
  {Zhang}}{2014}]{Deng}
{Deng} W.,  {Zhang} B.,  2014, \mn@doi [\apjl] {10.1088/2041-8205/783/2/L35},
  \href {http://adsabs.harvard.edu/abs/2014ApJ...783L..35D} {783, L35}

\bibitem[\protect\citeauthoryear{{Dolag}, {Gaensler}, {Beck}  \&
  {Beck}}{{Dolag} et~al.}{2015}]{Dolag}
{Dolag} K.,  {Gaensler} B.~M.,  {Beck} A.~M.,   {Beck} M.~C.,  2015, \mn@doi
  [\mnras] {10.1093/mnras/stv1190}, \href
  {http://adsabs.harvard.edu/abs/2015MNRAS.451.4277D} {451, 4277}

\bibitem[\protect\citeauthoryear{{Eftekhari} \& {Berger}}{{Eftekhari} \&
  {Berger}}{2017}]{Eftekhari2017}
{Eftekhari} T.,  {Berger} E.,  2017, \mn@doi [\apj] {10.3847/1538-4357/aa90b9},
  \href {http://adsabs.harvard.edu/abs/2017ApJ...849..162E} {849, 162}

\bibitem[\protect\citeauthoryear{{Fan}, {Narayanan}, {Strauss}, {White},
  {Becker}, {Pentericci}  \& {Rix}}{{Fan} et~al.}{2002}]{Fan}
{Fan} X.,  {Narayanan} V.~K.,  {Strauss} M.~A.,  {White} R.~L.,  {Becker}
  R.~H.,  {Pentericci} L.,   {Rix} H.-W.,  2002, \mn@doi [\aj]
  {10.1086/339030}, \href {http://adsabs.harvard.edu/abs/2002AJ....123.1247F}
  {123, 1247}

\bibitem[\protect\citeauthoryear{{Fukugita}, {Hogan}  \& {Peebles}}{{Fukugita}
  et~al.}{1998}]{Fukugita1998}
{Fukugita} M.,  {Hogan} C.~J.,   {Peebles} P.~J.~E.,  1998, \mn@doi [\apj]
  {10.1086/306025}, \href {http://adsabs.harvard.edu/abs/1998ApJ...503..518F}
  {503, 518}

\bibitem[\protect\citeauthoryear{{Gao}, {Li}  \& {Zhang}}{{Gao}
  et~al.}{2014}]{Gao}
{Gao} H.,  {Li} Z.,   {Zhang} B.,  2014, \mn@doi [\apj]
  {10.1088/0004-637X/788/2/189}, \href
  {http://adsabs.harvard.edu/abs/2014ApJ...788..189G} {788, 189}

\bibitem[\protect\citeauthoryear{{Hopkins} \& {Beacom}}{{Hopkins} \&
  {Beacom}}{2006}]{Hopkins}
{Hopkins} A.~M.,  {Beacom} J.~F.,  2006, \mn@doi [\apj] {10.1086/506610}, \href
  {http://adsabs.harvard.edu/abs/2006ApJ...651..142H} {651, 142}

\bibitem[\protect\citeauthoryear{{Ilbert} et~al.,}{{Ilbert}
  et~al.}{2006}]{Ilbert}
{Ilbert} O.,  et~al., 2006, \mn@doi [\aap] {10.1051/0004-6361:20065138}, \href
  {http://adsabs.harvard.edu/abs/2006A%26A...457..841I} {457, 841}

\bibitem[\protect\citeauthoryear{{Ioka}}{{Ioka}}{2003}]{Ioka}
{Ioka} K.,  2003, \mn@doi [\apjl] {10.1086/380598}, \href
  {http://adsabs.harvard.edu/abs/2003ApJ...598L..79I} {598, L79}

\bibitem[\protect\citeauthoryear{{Keane}}{{Keane}}{2018}]{Keane2018}
{Keane} E.~F.,  2018, \mn@doi [Nature Astronomy] {10.1038/s41550-018-0603-0},
  \href {http://adsabs.harvard.edu/abs/2018NatAs...2..865K} {2, 865}

\bibitem[\protect\citeauthoryear{{La Plante}, {Trac}, {Croft}  \& {Cen}}{{La
  Plante} et~al.}{2017}]{Plante}
{La Plante} P.,  {Trac} H.,  {Croft} R.,   {Cen} R.,  2017, preprint, \href
  {http://adsabs.harvard.edu/abs/2017arXiv171003286L} {} (\mn@eprint {arXiv}
  {1710.03286})

\bibitem[\protect\citeauthoryear{{Luo}, {Lee}, {Lorimer}  \& {Zhang}}{{Luo}
  et~al.}{2018}]{Luo}
{Luo} R.,  {Lee} K.,  {Lorimer} D.~R.,   {Zhang} B.,  2018, \mn@doi [\mnras]
  {10.1093/mnras/sty2364}, \href
  {http://adsabs.harvard.edu/abs/2018MNRAS.481.2320L} {481, 2320}

\bibitem[\protect\citeauthoryear{{Macquart} et~al.,}{{Macquart}
  et~al.}{2015}]{JPska}
{Macquart} J.~P.,  et~al., 2015, Advancing Astrophysics with the Square
  Kilometre Array (AASKA14), \href
  {http://adsabs.harvard.edu/abs/2015aska.confE..55M} {p.~55}

\bibitem[\protect\citeauthoryear{{Marcote} et~al.,}{{Marcote}
  et~al.}{2017}]{Marcote}
{Marcote} B.,  et~al., 2017, \mn@doi [\apjl] {10.3847/2041-8213/834/2/L8},
  \href {http://adsabs.harvard.edu/abs/2017ApJ...834L...8M} {834, L8}

\bibitem[\protect\citeauthoryear{{McQuinn}}{{McQuinn}}{2014}]{McQuinn}
{McQuinn} M.,  2014, \mn@doi [\apjl] {10.1088/2041-8205/780/2/L33}, \href
  {http://adsabs.harvard.edu/abs/2014ApJ...780L..33M} {780, L33}

\bibitem[\protect\citeauthoryear{{McQuinn}, {Lidz}, {Zaldarriaga}, {Hernquist},
  {Hopkins}, {Dutta}  \& {Faucher-Gigu{\`e}re}}{{McQuinn}
  et~al.}{2009}]{McQuinnHeII}
{McQuinn} M.,  {Lidz} A.,  {Zaldarriaga} M.,  {Hernquist} L.,  {Hopkins} P.~F.,
   {Dutta} S.,   {Faucher-Gigu{\`e}re} C.-A.,  2009, \mn@doi [\apj]
  {10.1088/0004-637X/694/2/842}, \href
  {http://adsabs.harvard.edu/abs/2009ApJ...694..842M} {694, 842}

\bibitem[\protect\citeauthoryear{{Parsons} et~al.,}{{Parsons}
  et~al.}{2014}]{Parsons}
{Parsons} A.~R.,  et~al., 2014, \mn@doi [\apj] {10.1088/0004-637X/788/2/106},
  \href {http://adsabs.harvard.edu/abs/2014ApJ...788..106P} {788, 106}

\bibitem[\protect\citeauthoryear{{Planck Collaboration} et~al.,}{{Planck
  Collaboration} et~al.}{2016}]{Planck}
{Planck Collaboration} et~al., 2016, \mn@doi [\aap]
  {10.1051/0004-6361/201525830}, \href
  {http://adsabs.harvard.edu/abs/2016A%26A...594A..13P} {594, A13}

\bibitem[\protect\citeauthoryear{{Shull}, {Smith}  \& {Danforth}}{{Shull}
  et~al.}{2012}]{Shull}
{Shull} J.~M.,  {Smith} B.~D.,   {Danforth} C.~W.,  2012, \mn@doi [\apj]
  {10.1088/0004-637X/759/1/23}, \href
  {http://adsabs.harvard.edu/abs/2012ApJ...759...23S} {759, 23}

\bibitem[\protect\citeauthoryear{{Singh} et~al.,}{{Singh} et~al.}{2018}]{Singh}
{Singh} S.,  et~al., 2018, \mn@doi [\apj] {10.3847/1538-4357/aabae1}, \href
  {http://adsabs.harvard.edu/abs/2018ApJ...858...54S} {858, 54}

\bibitem[\protect\citeauthoryear{{Sokasian}, {Abel}  \& {Hernquist}}{{Sokasian}
  et~al.}{2002}]{Sokasian}
{Sokasian} A.,  {Abel} T.,   {Hernquist} L.~E.,  2002, ArXiv Astrophysics
  e-prints, \href {http://adsabs.harvard.edu/abs/2002astro.ph..6428S} {}

\bibitem[\protect\citeauthoryear{{Stappers}}{{Stappers}}{2016}]{meertrap}
{Stappers} B.,  2016, in Proceedings of MeerKAT Science: On the Pathway to the
  SKA. 25-27 May, 2016 Stellenbosch, South Africa (MeerKAT2016). p.~10

\bibitem[\protect\citeauthoryear{{Straatman} et~al.,}{{Straatman}
  et~al.}{2016}]{Straatman}
{Straatman} C. M.~S.,  et~al., 2016, \mn@doi [\apj]
  {10.3847/0004-637X/830/1/51}, \href
  {https://ui.adsabs.harvard.edu/#abs/2016ApJ...830...51S} {830, 51}

\bibitem[\protect\citeauthoryear{Syphers et~al.,}{Syphers
  et~al.}{2009}]{SyphersHeII2}
Syphers D.,  et~al., 2009, The Astrophysical Journal, 690, 1181

\bibitem[\protect\citeauthoryear{Syphers, Anderson, Zheng, Meiksin, Schneider
  \& York}{Syphers et~al.}{2012}]{SyphersHeII}
Syphers D.,  Anderson S.~F.,  Zheng W.,  Meiksin A.,  Schneider D.~P.,   York
  D.~G.,  2012, The Astronomical Journal, 143, 100

\bibitem[\protect\citeauthoryear{{Tendulkar} et~al.,}{{Tendulkar}
  et~al.}{2017}]{SriHarsh}
{Tendulkar} S.~P.,  et~al., 2017, \mn@doi [\apjl] {10.3847/2041-8213/834/2/L7},
  \href {http://adsabs.harvard.edu/abs/2017ApJ...834L...7T} {834, L7}

\bibitem[\protect\citeauthoryear{{Tingay} et~al.,}{{Tingay}
  et~al.}{2013}]{Tingay}
{Tingay} S.~J.,  et~al., 2013, \mn@doi [\pasa] {10.1017/pasa.2012.007}, \href
  {http://adsabs.harvard.edu/abs/2013PASA...30....7T} {30, e007}

\bibitem[\protect\citeauthoryear{{Walker}, {Ma}  \& {Breton}}{{Walker}
  et~al.}{2018}]{CWalker}
{Walker} C.~R.~H.,  {Ma} Y.-Z.,   {Breton} R.~P.,  2018, arXiv e-prints, \href
  {http://adsabs.harvard.edu/abs/2018arXiv180401548W} {}

\bibitem[\protect\citeauthoryear{{Xu} \& {Han}}{{Xu} \& {Han}}{2015}]{Xu}
{Xu} J.,  {Han} J.~L.,  2015, preprint, \href
  {http://adsabs.harvard.edu/abs/2015arXiv150400200X} {} (\mn@eprint {arXiv}
  {1504.00200})

\bibitem[\protect\citeauthoryear{{Yao}, {Manchester}  \& {Wang}}{{Yao}
  et~al.}{2017}]{Yao}
{Yao} J.~M.,  {Manchester} R.~N.,   {Wang} N.,  2017, \mn@doi [\apj]
  {10.3847/1538-4357/835/1/29}, \href
  {http://adsabs.harvard.edu/abs/2017ApJ...835...29Y} {835, 29}

\bibitem[\protect\citeauthoryear{{Young}}{{Young}}{1976}]{Young}
{Young} P.~J.,  1976, \mn@doi [\aj] {10.1086/111959}, \href
  {http://adsabs.harvard.edu/abs/1976AJ.....81..807Y} {81, 807}

\bibitem[\protect\citeauthoryear{{Zaroubi}}{{Zaroubi}}{2013}]{Zaroubi}
{Zaroubi} S.,  2013, in {Wiklind} T.,  {Mobasher} B.,   {Bromm} V.,  eds,
  Astrophysics and Space Science Library Vol. 396, The First Galaxies. p.~45
  (\mn@eprint {arXiv} {1206.0267}), \mn@doi{10.1007/978-3-642-32362-1_2}

\bibitem[\protect\citeauthoryear{{Zhang}}{{Zhang}}{2018}]{BZhang}
{Zhang} B.,  2018, \mn@doi [\apjl] {10.3847/2041-8213/aae8e3}, \href
  {http://adsabs.harvard.edu/abs/2018ApJ...867L..21Z} {867, L21}

\bibitem[\protect\citeauthoryear{{Zheng}, {Ofek}, {Kulkarni}, {Neill}  \&
  {Juric}}{{Zheng} et~al.}{2014}]{Zheng}
{Zheng} Z.,  {Ofek} E.~O.,  {Kulkarni} S.~R.,  {Neill} J.~D.,   {Juric} M.,
  2014, \mn@doi [\apj] {10.1088/0004-637X/797/1/71}, \href
  {http://adsabs.harvard.edu/abs/2014ApJ...797...71Z} {797, 71}

\bibitem[\protect\citeauthoryear{{Zhou}, {Li}, {Wang}, {Fan}  \& {Wei}}{{Zhou}
  et~al.}{2014}]{Zhou}
{Zhou} B.,  {Li} X.,  {Wang} T.,  {Fan} Y.-Z.,   {Wei} D.-M.,  2014, \mn@doi
  [\prd] {10.1103/PhysRevD.89.107303}, \href
  {http://adsabs.harvard.edu/abs/2014PhRvD..89j7303Z} {89, 107303}

\bibitem[\protect\citeauthoryear{{van Haarlem} et~al.,}{{van Haarlem}
  et~al.}{2013}]{vanHaarlem}
{van Haarlem} M.~P.,  et~al., 2013, \mn@doi [\aap]
  {10.1051/0004-6361/201220873}, \href
  {http://adsabs.harvard.edu/abs/2013A%26A...556A...2V} {556, A2}

\makeatother
\end{thebibliography}

%%%%%%%%%%%%%%%%%%%%%%%%%%%%%%%%%%%%%%%%%%%%%%%%%%

% Don't change these lines
\bsp	% typesetting comment
\label{lastpage}
\end{document}